# QUANTUM MECHANICAL DESCRIPTION OF BELL'S EXPERIMENT ASSUMES LOCALITY

**Alejandro A. Hnilo**

*CEILAP, Centro de Investigaciones en Láseres y Aplicaciones, (MINDEF-CONICET);*
*J.B. de La Salle 4397, (1603) Villa Martelli, Argentina.*
*email: alex.hnilo@gmail.com*

*Abstract.*

In the usual derivation of Bell's inequalities, Locality is assumed to justify statistical independence of the outcomes of observations performed in remote stations; as the inequalities are observed to be violated, it is concluded that these outcomes are statistically dependent, and hence, that non-Local effects exist. Here it is shown that the simplest description of Bell's experiment according to the canon of von Neumann's theory of measurement assumes statistical independence of those outcomes. In other words, the straightforward description of Bell' experiment according to Quantum Mechanics (which, of course, violates Bell's inequalities) does not assume non-Locality. Summing up this and other related results it is concluded that, within the framework of Quantum Mechanics, there is no reason to believe in the existence of non-Local effects. In addition to its foundational significance, this conclusion has practical impact in the fields of quantum-certified and device-independent randomness generation and on the vulnerability of Quantum Key Distribution schemes, which are based on the existence of quantum non-Locality as an axiom.

Important and plentiful research activity is based on the idea that sequences of outcomes of successive measurements performed on a set of identically prepared spatially extended entangled states (or Bell's experiment, see Figure 1) are necessarily random. It is even claimed this to be the *only* source of "true" randomness. This idea is based on the result that non-Local effects (which are assumed exist in such experiment, as an axiom) and the impossibility of faster-than-light signaling imply that said sequences must be fully unpredictable [1-3]. On the other hand, it has been demonstrated that non-Local effects appear when the Bell's experiment is described within the realm of classical hidden variable theories, but that they vanish when the description is within the realm of Quantum Mechanics (QM) [4]. It has been also demonstrated that the violation of Bell's inequalities has nothing to do with non-Local effects, but that it is the mere consequence of non-commutativity of *local* observables. A similar result had been conjectured long ago in terms of wave-particle dualism [5]. Besides, it has been shown that it is possible to remove quantum non-Locality by an appropriate interpretation of wavefunction's properties [6]. In spite of these and other arguments [7] against quantum non-Locality, as A.Khrennikov has warned: "*It is clear that to get rid of nonlocality from quantum theory is not a simple task. The present note is just a step towards the common acceptance of the local interpretation of QM*". This paper can be regarded as another step in the same direction. Note that deciding the existence or non-existence of non-Local effects has an importance not only foundational but also practical, for it affects a much sought definition of "random" [8], the bases of quantum-certified random number generators, and the claimed invulnerability of quantum key distribution devices.

This paper describes Bell's experiment according to von Neumann's canonical description of quantum measurement. As it will be shown, the description does not assume non-Local effects but, on the contrary, it assumes statistical independence of the outcomes recorded in the remote stations. Recall that what is needed in the usual derivation of Bell's inequalities is this statistical independence, which is a consequence of assuming Locality (it is then usually argued that violation of Bell's inequalities implies statistical *dependence*, and hence, the existence of non-Local effects). However, von Neumann's description assumes statistical independence and still violates Bell's inequalities. Non Local effects are, in consequence, unnecessary to explain said violation within the quantum theory.

Von Neumann's theory of measurement is based on the definition of an appropriate interaction between the quantum system and a classical system, called the "pointer". The latter is eventually observed. This theory [9] is closely followed here, but it is applied to a pair of entangled quantum particles instead that to a single one. Consider then a typical Bell's experiment (Fig.1). Assume source S emits biphotons in the fully symmetrical polarization-entangled Bell state: $|\Phi^+\rangle$ =

$(1/\surd 2)(|x_A,x_B\rangle + |y_A,y_B\rangle)$. At remote stations Alice and Bob the bipohotons are projected into the polarizers' bases $\{|+A\rangle,|-A\rangle\}$ and $\{|+B\rangle,|-B\rangle\}$. Eigenvalue +1 (-1) corresponds to a photon detected after the transmitted (reflected) polarizer's output. These bases are related with the linear {x,y} polarization ones as:

$$|x_A\rangle = |+A\rangle \cos(\alpha) - |-A\rangle \sin(\alpha), \quad |y_A\rangle = |+A\rangle \sin(\alpha) + |-A\rangle \cos(\alpha) \tag{1}$$
$$|x_B\rangle = |+B\rangle \cos(\beta) - |-B\rangle \sin(\beta), \quad |y_B\rangle = |+B\rangle \sin(\beta) + |-B\rangle \cos(\beta)$$

where $\alpha$ ($\beta$) is the orientation angle of the polarization analyzer in station A (B) with respect to the {x,y} axes. The operators for the observables in each station are then (note the bold typing for operators):

$$\mathbf{A} = (+1)\, |+A\rangle\langle +A| + (-1)\, |-A\rangle\langle -A|, \quad \mathbf{B} = (+1)\, |+B\rangle\langle +B| + (-1)\, |-B\rangle\langle -B| \tag{2}$$

Following von Neumann's theory, the interaction Hamiltonian with the classical apparatus in station A is assumed $\mathbf{H_A} = \lambda \mathbf{A}\times\mathbf{P_A}$, where $\lambda$ measures the strength of the interaction, and $\mathbf{P_A}$ is the linear momentum operator of the pointer in that station (also $\mathbf{H_B} = \lambda \mathbf{B}\times\mathbf{P_B}$, the interaction strength is assumed the same in both stations for simplicity). The evolution of the system "entangled state plus apparatuses" is:

$$|\psi_{final}\rangle = \mathbf{U}(t)\, |\psi_{initial}\rangle = \mathbf{exp[}(-2\pi i t/h)\, \mathbf{(H_A + H_B)]}\, \{(1/\surd 2)(|x_A,x_B\rangle + |y_A,y_B\rangle)\}|\psi^0{}_A\rangle|\psi^0{}_B\rangle \tag{3}$$

where $|\psi^0{}_A\rangle$, $|\psi^0{}_B\rangle$ are the pointers' initial states (see Fig.1, left).

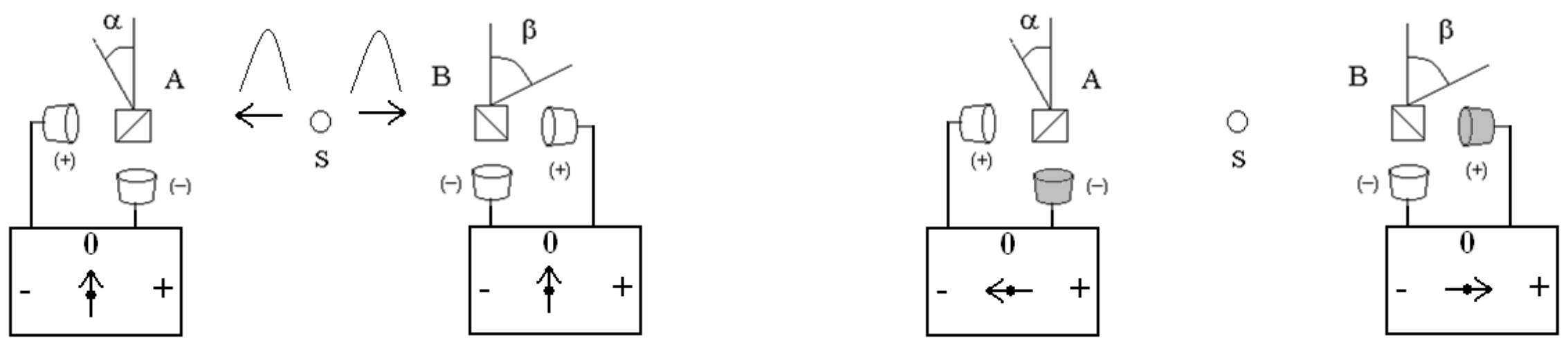

Figure 1: Sketch of a Bell's experiment. *Left*: biphotons in the fully symmetrical Bell state $|\Phi^+\rangle$ are emitted by source S and propagate to stations A and B, where pointers of classical apparatuses are in their initial states or positions. After passing through polarization analyzers set at angles $\{\alpha,\beta\}$, the biphotons interact with the apparatuses, as described by von Neumann's theory of measurement. *Right*: after the interaction is completed the pointer in station A has evolved (in this example) into the classical state "-", the one in B into the "+". The outcome "-+" is therefore observed. Probabilities of each outcome {++, +-, -+, --} are correctly given by eq.(9), which is derived by assuming statistical independence of the outcomes in A and B (here, the QM-language equivalent of Locality).

Using the seemingly trivial equality:

$$\mathbf{exp[}(-2\pi it/h)\ \mathbf{(H_A + H_B)] = exp[}(-2\pi it/h)\ \mathbf{H_A]} \times \mathbf{exp[}(-2\pi it/h)\ \mathbf{H_B]} \quad (4)$$

replacing $\mathbf{H_A}$, $\mathbf{H_B}$, using eqs.(1,2) and the property that for operators **A**, **B** with eigenvalues *a*, *b*:

$$\mathbf{exp(}i\ \mathbf{A{\times}B)} \approx \mathbf{A} \times \mathbf{exp(}ia\mathbf{B)} \quad (5)$$

then:

$|\psi_{final}\rangle = (1/\surd 2)\{\cos(\alpha).\cos(\beta).|+A\rangle|+B\rangle\ \mathbf{exp}(-i\varepsilon\mathbf{P_A})\ |\psi^0{}_A\rangle\ \mathbf{exp}(-i\varepsilon\mathbf{P_B})\ |\psi^0{}_B\rangle +$

$\sin(\alpha).\sin(\beta).|+A\rangle|+B\rangle\ \mathbf{exp}(-i\varepsilon\mathbf{P_A})\ |\psi^0{}_A\rangle\ \mathbf{exp}(-i\varepsilon\mathbf{P_B})\ |\psi^0{}_B\rangle +$

$\cos(\alpha).[-\sin(\beta)].|+A\rangle|-B\rangle\ \mathbf{exp}(-i\varepsilon\mathbf{P_A})\ |\psi^0{}_A\rangle\ \mathbf{exp}(i\varepsilon\mathbf{P_B})\ |\psi^0{}_B\rangle +$

$\sin(\alpha).\cos(\beta).|+A\rangle|-B\rangle\ \mathbf{exp}(-i\varepsilon\mathbf{P_A})\ |\psi^0{}_A\rangle\ \mathbf{exp}(i\varepsilon\mathbf{P_B})\ |\psi^0{}_B\rangle +$

$[-\sin(\alpha)].\cos(\beta).|-A\rangle|+B\rangle\ \mathbf{exp}(i\varepsilon\mathbf{P_A})\ |\psi^0{}_A\rangle\ \mathbf{exp}(-i\varepsilon\mathbf{P_B})\ |\psi^0{}_B\rangle +$

$\cos(\alpha).\cos(\beta).|-A\rangle|+B\rangle\ \mathbf{exp}(i\varepsilon\mathbf{P_A})\ |\psi^0{}_A\rangle\ \mathbf{exp}(-i\varepsilon\mathbf{P_B})\ |\psi^0{}_B\rangle +$

$[-\sin(\alpha)].[-\sin(\beta)].|-A\rangle|-B\rangle\ \mathbf{exp}(i\varepsilon\mathbf{P_A})\ |\psi^0{}_A\rangle\ \mathbf{exp}(i\varepsilon\mathbf{P_B})\ |\psi^0{}_B\rangle +$

$\cos(\alpha).\cos(\beta).|-A\rangle|-B\rangle\ \mathbf{exp}(i\varepsilon\mathbf{P_A})\ |\psi^0{}_A\rangle\ \mathbf{exp}(i\varepsilon\mathbf{P_B})\ |\psi^0{}_B\rangle\}$ (6)

where $\varepsilon \equiv 2\pi t\lambda/h$. Now assume that interaction time is short enough, and interaction strength large enough, so that $t\lambda \approx$ constant ("sudden interaction" approximation [9]), therefore $\varepsilon \approx$ constant. The linear momentum operator produces the pointer's displacement to the state:

$$\mathbf{exp}(-i\varepsilon\mathbf{P_A})\ |\psi^0{}_A\rangle = |\psi^+{}_A\rangle \quad (7)$$

what means "state of the classical apparatus in A with its pointer in the position +". The pointer no longer moves. It indicates that a single detection has occurred in the output "transmitted" of the polarizer in station A. The other operators in eq.(6) produce analogous displacements to the pointers' final states: $|\psi^+{}_B\rangle$, $|\psi^-{}_A\rangle$, $|\psi^-{}_B\rangle$. In consequence:

$|\psi_{final}\rangle = (1/\surd 2)\{\ |+A\rangle|+B\rangle|\psi^+{}_A\rangle|\psi^+{}_B\rangle.\ [\cos(\alpha).\cos(\beta)+\sin(\alpha).\sin(\beta)] +$

$|+A\rangle|-B\rangle|\psi^+{}_A\rangle|\psi\text{-}_B\rangle.\ [-\cos(\alpha).\sin(\beta)+\sin(\alpha).\cos(\beta)] +$

$|-A\rangle|+B\rangle|\psi^-{}_A\rangle|\psi^+{}_B\rangle.\ [-\sin(\alpha).\cos(\beta)+\cos(\alpha).\sin(\beta)] +$

$|-A\rangle|-B\rangle|\psi^-{}_A\rangle|\psi^-{}_B\rangle.\ [\sin(\alpha).\sin(\beta)+\cos(\alpha).\cos(\beta)]\ \}$ (8)

Note that the states of the quantum system and the classical pointers are entangled now. But the observer has no access to the quantum system, only to the pointers. What the observer sees is then described by a reduced density matrix, obtained after tracing out the states $\{|+A\rangle,|+B\rangle,|-A\rangle,|-B\rangle\}$ of the quantum system:

$$\rho_{reduced} = \Sigma \langle \textit{quantum system}|\psi_{final}\rangle\langle\psi_{final}| \textit{ quantum system}\rangle = \tfrac{1}{2} \{\cos^2(\alpha-\beta).|\psi^+_A\rangle|\psi^+_B\rangle\langle\psi^+_A|\langle\psi^+_B| + \sin^2(\alpha-\beta).|\psi^+_A\rangle|\psi^-_B\rangle\langle\psi^+_A|\langle\psi^-_B| + \sin^2(\alpha-\beta).|\psi^-_A\rangle|\psi^+_B\rangle\langle\psi^-_A|\langle\psi^+_B| + \cos^2(\alpha-\beta).|\psi^-_A\rangle|\psi^-_B\rangle\langle\psi^-_A|\langle\psi^-_B|\} \quad (9)$$

which is, as expected, a diagonal density matrix with the correct probabilities for the four possible outcomes in Fig.1 on the right.

Statistical independence is assumed in the (apparently naïve) eq.(4), valid if $[\mathbf{H_A}, \mathbf{H_B}] = \mathbf{0}$. In fact, if these Hamiltonians did not commute, the results of observations could be different in different reference frames and allow faster than light signaling. The commutation of the Hamiltonians is therefore equivalent, in this quantum description, to the condition of Locality in the usual (classical) derivation of Bell's inequalities. Of course, it is still possible that some ad-hoc election of $\mathbf{H_A}, \mathbf{H_B}$ may lead to the correct result eq.(9) even if $[\mathbf{H_A}, \mathbf{H_B}] \neq \mathbf{0}$, but the simplest, natural way to derive eq.(9) assumes $[\mathbf{H_A}, \mathbf{H_B}] = \mathbf{0}$. Strictly speaking quantum non-Locality is not refuted, but it is demonstrated unnecessary. By Occam's razor, it must be discarded.

Summing up the arguments in [4-8] and here, it is concluded that there is no reason to believe on the existence of quantum non-Local effects. It is pertinent recalling that Bohr's immediate reaction to EPR's paper was that Realism (not Locality) was false in QM. Bohr's point of view is hence strengthened here. Nevertheless, as stated by A.Khrennikov, it is to be expected that many papers and discussions will be necessary before the apparently fruitful and deeply rooted belief on the existence of quantum non-Locality is abandoned. An important consequence of the conclusion above is that the ideas of quantum-certified and device-independent randomness, which are derived from the hypothesis of quantum non-Locality as an axiom [1], are weakened. This has practical impact on important developments in the field of quantum Random Number Generators and the security of Quantum Key Distribution schemes using entangled states.

**Acknowledgements.**

This work received support from grants N62909-18-1-2021 Office of Naval Research Global (USA), and PIP 2017-027 CONICET (Argentina).

**References.**

[1] S.Popescu and D.Rohrlich, "Quantum nonlocality as an axiom", *Found. Phys.* **24** p.379 (1994).

[2] C.Calude and K.Svozil, “Quantum randomness and value indefiniteness,” *Advanced Science Lett.* **1**, p.165 (2008).
[3] S.Pironio *et al.*, “Random numbers certified by Bell’s theorem”, *Nature* **23**, p.1021 (2010).
[4] A.Khrennikov, “Get Rid of Nonlocality from Quantum Physics”, *Entropy* **2019**, 21, 806; doi:10.3390/e21080806.
[5] A.Garuccio and F.Selleri, “Nonlocal interactions and Bell’s inequality”, *Nuovo Cim.* **36B** p.176 (1976).
[6] H.Zwirn, “Non Locality vs. modified Realism” *Found. Phys.* **50** p.1 (2020).
[7] M.Kupczynski, “Closing the Door on Quantum Nonlocality”, *Entropy* **2018**, 20, 877; doi:10.3390/e20110877.
[8] M.Agüero *et al.*, “Time evolution of randomness in Bell’s experiment indicates “Realism” is false”, *arxiv*:2504.03921, see the Supplementary Material section 3.1.
[9] J.Preskill, *Quantum Computation Lecture Notes*, Caltech Ph219/CS219, (2013-2014).